\input epsf
\magnification=1200
\null
$\qquad\qquad\qquad\qquad\qquad\qquad\qquad\qquad\qquad\qquad\qquad\qquad\qquad\qquad$
{May 16, 2010}
\bigskip\bigskip\bigskip\bigskip\noindent

\centerline{PHASE TRANSITIONS WITH FOUR-SPIN INTERACTIONS.}
\bigskip\bigskip
\centerline{by Joel L. Lebowitz\footnote{*}{Math. Dept., Rutgers University, and IHES, 91440 Bures sur Yvette, France.}\footnote{$\dagger$}{Physics Dept., Rutgers University.}\footnote{}{emails: $<$lebowitz@math.rutgers.edu$>$, $<$ruelle@ihes.fr$>$.} and David Ruelle$^*$.}
\bigskip\bigskip\bigskip\bigskip
$\qquad\qquad\qquad\qquad\qquad\qquad\qquad\qquad\qquad\qquad\qquad\qquad$
{\sl In memory of Julius Borcea}
\bigskip\bigskip\bigskip\bigskip\noindent
	{\leftskip=2cm\rightskip=2cm\sl Abstract.  Using an extended Lee-Yang theorem and GKS correlation inequalities, we prove, for a class of ferromagnetic multi-spin interactions, that they will have a phase transition(and spontaneous magnetization) if, and only if, the external field $h=0$ (and the temperature is low enough).  We also show the absence of phase transitions for some nonferromagnetic interactions.  The FKG inequalities are shown to hold for a larger class of multi-spin interactions.\par}
\vfill\eject
	{\bf 1. Introduction.}
\medskip
	The mathematically best understood systems exhibiting phase transitions are Ising models with ferromagnetic interactions on $\Lambda\subset{\bf Z}^d$.  In addition to the special exactly solvable examples, e.g., the nearest neighbor Ising model on ${\bf Z}^2$, ``almost" everything is known qualitatively about the phase diagram and correlation functions of such systems.  This is due to the Lee-Yang theorem [LY], the Asano-Ruelle lemma, correlation inequalities, and low-temperature analysis based on the (broken) symmetry of the ground state\footnote{(*)}{For a general introduction to rigorous equilibrium statistical mechanics, in particular lattice spin systems, see [R1], [I], [Sin], [Sim].  The results needed here about Lee-Yang and Asano-Ruelle are contained in [R3] and summarized in Appendix A of the present paper, but there is a vast literature on the subject (C.M. Newman, E.H. Lieb, A.D. Sokal, J. Chayes, etc.); the reader may consult [BB1] and [BB2] for a different approach and a large list of references.  For correlation inequalities see Section 8.  For low-temperature analysis see [Sin], [Sl] (Pirogov-Sinai theory) and also [HS].}.  In this note we extend some of these results to a class of systems with both pair and four spin interactions.  In particular we show that the thermodynamic and correlation functions are analytic if the external magnetic field $h$ does not vanish.  Conversely, they will have, when the interactions are ferromagnetic, a phase transition when $h=0$ and the temperature is low enough.  There is also a class of interactions for which there are no phase transitions at any temperature.
\medskip
	Our results are obtained by the use of an extended Lee-Yang theorem [R3] in combination with GKS inequalities.  We also obtain FKG inequalities for systems with multispin interactions.  Finally, we prove Lee-Yang type ferromagnetic behavior for a large class of systems with multispin interactions at sufficiently low temperatures.
\bigskip
	{\bf 2. The model.}
\medskip
	In this note we study classical spin systems such that a configuration $\underline\sigma=(\sigma_x)_{x\in\Lambda}$ of spins $\sigma_x=\pm1$ in a finite region $\Lambda$ has energy $U(\underline\sigma)-\sum_xh_x\sigma_x$, where $h_x$ is the magnetic field at $x$.  It is convenient to use instead of the configuration $\underline\sigma$ the set $X=\{x:\sigma_x=+1\}$ of ``occupied'' sites and to consider the partition function
$$	Z_\Lambda(\underline z)
	=\sum_{X\subset\Lambda}e^{-\beta U_X}\prod_{x\in X}e^{2\beta h_x}
	=\sum_{X\subset\Lambda}E_Xz^X\eqno{(2.1)}   $$
where we have written $U(\underline\sigma)=U_X$, $E_X=e^{-\beta U_X}$, $z_x=e^{2\beta h_x}$, $\underline z=(z_x)_{x\in\Lambda}$, $z^X=\prod_{x\in X}z_x$.  The {\it spin-flip symmetry} $U(\underline\sigma)=U(-\underline\sigma)$ is now expressed by $E_X=E_{\Lambda\backslash X}$.
\medskip
	It is known [R3] that in the $2^{|\Lambda|-1}$-dimensional space of partition functions $Z_\Lambda$ satisfying the spin-flip symmetry there is a nonempty open set such that the {\it Lee-Yang property} is satisfied: $Z_\Lambda(\underline z)\ne0$ if $|z_x|<1$ for all $x$ \footnote{(**)}{The spin-flip  symmetry implies that $Z_\Lambda(\underline z)\ne0$ also if $|z_x|>1$ for all $x$, and therefore we have the {\it Circle Theorem}: all zeros of $Z_\Lambda(z,\ldots,z)$ are on the unit circle $\{z:|z|=1\}$.}.  Furthermore if the Lee-Yang property is satisfied at high temperature, it is satisfied at all temperatures, and the energy is defined by a ferromagnetic pair interaction between spins (see [R3] Theorem 9).
\medskip
	In what follows we shall discuss a specific class of models with 4-spin interactions, for which the Lee-Yang property is violated at high temperatures, but such that, at any temperature for which the zeros of $Z_\Lambda(z,\ldots,z)$ come close to the positive real axis, the Lee-Yang property holds.  In the thermodynamic limit therefore, a phase transition can only occur at zero magnetic field, and examples exist where spontaneous magnetization does indeed occur.  The methods used for the proofs will be generalized Lee-Yang theory (with the Asano-Ruelle lemma), and correlation inequalities.
\medskip
	Let us expand on what is meant here by absence of phase transition.  Consider a bounded finite-range perturbation $\lambda V_X$ of the energy $U_X$ in $(2.2)$ [i.e., $V_X=0$ unless diameter $X<A$ and $|V_X|<B$ for all $X$, translation invariance is not required].  We write
$$	Z_\Lambda^\lambda(\underline z)
	=\sum_{X\subset\Lambda}e^{-\beta(U_X+\lambda V_X)}z^X   $$
Then, for small $\lambda$, the Asano-Ruelle lemma gives regions free of zeros for $Z_\Lambda^\lambda(\underline z)$ [see in particular [LR] in this respect].  Thus, if suitable translation invariance or periodicity is assumed for $V$, the free energy is analytic in $\lambda$, except in the ferromagnetic case, close to $h=0$, at low temperature.
\medskip
	A concrete version of the models to be discussed is obtained from a planar square Ising model by adding, on alternate {\it distinguished} squares, diagonal interactions of the same strength as the interactions on the sides, and a 4-spin interaction (marked by a circle in the figure below).  Note that only two distinguished squares meet at each lattice vertex.  This will be important for the proof of (i) in Section 7.  The proof would fail if there were 4-spin interactions on each square, not just alternate squares.
\smallskip
\centerline{\epsfbox{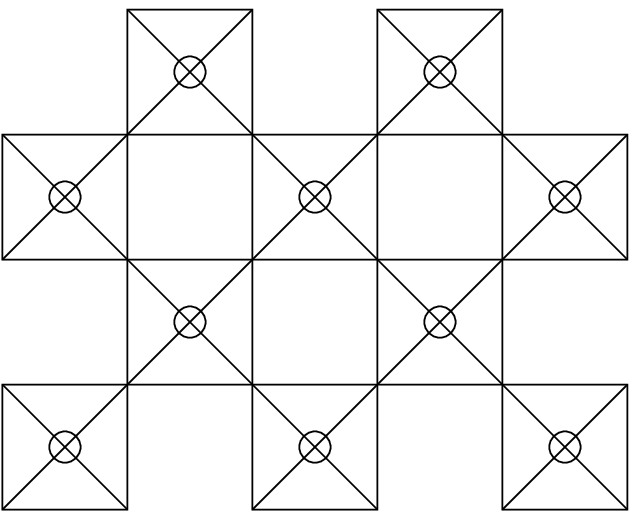}}
\noindent
The energy of a spin configuration $\underline\sigma$ is thus a sum over distinguished sqares $\alpha$ with four spins $\sigma_1^\alpha,\sigma_2^\alpha,\sigma_3^\alpha,\sigma_4^\alpha$, of contributions of the form
$$	U^\alpha=-J_2\sum_{i<j}\sigma_i^\alpha\sigma_j^\alpha
-J_4\sigma_1^\alpha\sigma_2^\alpha\sigma_3^\alpha\sigma_4^\alpha\eqno{(2.2)}$$
\indent
	The partition function $Z_\Lambda$ of a union $\Lambda$ of distinguished squares is obtained from the product $\Pi$ of the partition functions $Z^\alpha(z_1^\alpha,\ldots,z_4^\alpha)$ of the component distinguished squares $\alpha$ by a succession of steps $\Pi=\Pi_{x_0}\to\cdots\to\Pi_x\to\cdots\to Z_\Lambda$.  Each step involves a vertex $x$ of $\Lambda$ as follows:

\noindent
1) If $x$ coincides with the vertex $i$ of a single distinguished square $\alpha$, replace in $\Pi_x$ the corresponding $z_i^\alpha$ by $z_x$

\noindent
2) If $x$ belongs to the vertices $i,j$ of two squares $\alpha$ and $\beta$, write $\Pi_x=A+Bz_i^\alpha+Cz_j^\beta+Dz_i^\alpha z_j^\beta$, and replace $\Pi_x$ by $A+Dz_x$ (Asano contraction).  

\noindent
It is easy to see that, when all variables $z_i^\alpha$ have been eliminated in favor of the $z_x$ (the order in which this is done is unimportant) one obtains $Z_\Lambda$.
\medskip
	The interest of Asano contractions is that if one knows that $A+Bu+Cv+Duv\ne0$ when $u\notin K$, $v\notin L$, where $K,L$ are closed subsets of ${\bf C}$ disjoint from $0$, then $A+Dz\ne0$ when $z\notin-KL=\{-uv:u\in K,v\in L\}$ (see Lemma 1 of Appendix).  Also, because the partition function $Z^\alpha$ of a distinguished square is symmetric, one knows that $Z(u_1,u_2,u_3,u_4)\ne0$ if $u_1,u_2,u_3,u_4\notin K$, where K is any closed circular region containing the 4 zeros of $Z^\alpha(u,u,u,u)$ (see Lemma 2 of Appendix: Grace's theorem).
\bigskip
	{\bf 3. Main results.}
\medskip
	For the model described above, where $J_2\in{\bf R}$ is arbitrary and $J_4\ge0$, either or both of the following properties hold:
	
\noindent
(i) there is a neighborhood (independent of $\Lambda$) of the positive real axis $\{z\in {\bf C}: {\rm Im}z=0 \hbox{ and Re}z\ge0\}$ which is free of zeros of $Z_\Lambda(z,\ldots,z)$,

\noindent
(ii) $Z_\Lambda$ satisfies the Lee-Yang property.

\noindent
Therefore there can be no phase transition except at zero magnetic field.  Furthermore, if $J_2>0$, $J_4\ge0$, a phase transition actually occurs at low temperature.  If $J_4>0$, and $\Lambda$ contains at least one distinguished square, the Lee-Yang property for $Z_\Lambda$ is actually violated at sufficiently high temperature.  We also find that there is no phase transition when $\beta$ and $J_4\ge0$ satisfy $\beta J_2\le(\log 3)/8$; this condition always holds if $J_2\le0$.
\bigskip
	{\bf 4. Remark: other models.}
\medskip
	In order to make $Z^\alpha(z_1^\alpha,\ldots,z_4^\alpha)$ a symmetric function, we have made the contrived assumption that the diagonal interactions in a distinguished square have the same strength as the other interactions.  One can actually still obtain the same results as above under more natural conditions\footnote{(*)}{Using the notation of Section 5 below, we replace the $K_2$ diagonal interactions by $0$ (this leaves $J_2=K_4/8$), and we assume $K_2,K_4\ge0$.}but the proof is more acrobatic, and left to Appendix B.  One can also, instead of a planar system, consider a 3-dimensional model where the spins sit at the middle of the bonds of a diamond lattice.  A distinguished square is now replaced by a regular tetrahedron, the four spins at the vertices are in a completely symmetric situation, and the above results remain precisely the same in this new situation.
\medskip
	Our general strategy, which could certainly be used for other multiple spin interactions is as follows: (i) using the properties of Asano contractions show that, at high temperature, the partition function $Z_\Lambda$ has no zeros close to the real axis, (ii) show that at low temperature $Z_\Lambda$ satisfies the Lee-Yang property, (iii) use correlation inequalities to prove that a ferromagnetic phase transition is actually present at low temperature.
\bigskip
	{\bf 5. Partition function of a distinguished square.}
\medskip
	Let the energy of a set of 4 spins be given by the symmetric expression
$$	U(\sigma_1,\cdots,\sigma_4)=-K_0-K_2\sum_{1\le i<j\le4}
	[{1+\sigma_i\over2}\cdot{1+\sigma_j\over2}
	+{1-\sigma_i\over2}\cdot{1-\sigma_j\over2}]   $$
$$	-K_4[{1+\sigma_1\over2}\cdot{1+\sigma_2\over2}
	\cdot{1+\sigma_3\over2}\cdot{1+\sigma_4\over2}
	+{1-\sigma_1\over2}\cdot{1-\sigma_2\over2}
	\cdot{1-\sigma_3\over2}\cdot{1-\sigma_4\over2}]\eqno{(5.1)}   $$
$$	=-K_0-{1\over2}K_2\sum_{i<j}[1+\sigma_i\sigma_j]
-{1\over8}K_4[1+\sum_{i<j}\sigma_i\sigma_j+\sigma_1\sigma_2\sigma_3\sigma_4]  $$
$$	=-J_0-J_2\sum_{i<j}\sigma_i\sigma_j-J_4\sigma_1\sigma_2\sigma_3\sigma_4   $$
with $J_4=K_4/8$, $J_2=K_4/8+K_2/2$, $J_0=K_4/8+K_2/2+K_0$.  ($K_0$, $J_0$ are unimportant additive constants.)  It will be convenient to take $K_0=-3K_2$.
\medskip
	We may write
$$	Z(z_1,z_2,z_3,z_4)=\sum_{\sigma_1=\pm1}\cdots\sum_{\sigma_4=\pm1}
	\exp[-\beta(U(\sigma_1,\cdots,\sigma_4)+\sum_i h_i(1+\sigma_i))]   $$
$$	=bz_1z_2z_3z_4+(z_2z_3z_4+z_1z_3z_4+z_1z_2z_4+z_1z_2z_3)   $$
$$	+a(z_1z_2+z_1z_3+z_1z_4+z_2z_3+z_2z_4+z_3z_4)+(z_1+z_2+z_3+z_4)+b   $$
where $z_i=e^{2\beta h_i}$, $a=e^{-\beta K_2}$, $b=e^{\beta(K_4+3K_2)}$.  The symmetric polynomial $Z$ is entirely determined by
$$	P_{ab}(z)=Z(z,\cdots,z)=bz^4+4z^3+6az^2+4z+b   $$
and one can apply Grace's theorem to obtain properties of the zeros of $Z(z_1,\cdots,z_4)$ if one knows the zeros of $P_{ab}$.
\bigskip
	{\bf 6. Proposition} (the zeros of the partition function $P_{ab}$).
\medskip
	{\sl Let $a,b$ be real $>0$.

\noindent	
$\bullet$ Condition (i): the zeros of $P_{ab}$ are in the open left-hand plane $\{z:{\rm Re}z<0\}$ is implied by $3a-b>0$ (i.e., $e^{8\beta J_2}<3$).

\noindent
$\bullet$ Condition (ii): the zeros of $P_{ab}$ are on the unit circle $\{z:|z|=1\}$ is implied by ($1\le b$, and $3a\le b+2/b$, and $4\le 3a+b$).

\noindent
$\bullet$ The condition $4\le3a+b$ implies (i) or (ii).

\noindent
This condition, i.e., $4e^{2\beta(J_2-J_4)}\le3+e^{8\beta J_2}$ is satisfied in particular by $J_4\ge0$, all $J_2$.}
\medskip
	We have $P_{11}(z)=(z+1)^4$, so that (i) holds for $a=b=1$.  Since $b>0$, the zeros of $P_{ab}$ will remain in the left-hand plane until they cross the imaginary axis $\{z:{\rm Re}z=0\}$.  This will happen when $P_{ab}(iy)=0$ with real $y$, i.e., when $by^4-4iy^3-6ay^2+4iy+b=0$, or
$$ by^4-6ay^2+b=0\qquad{\rm and}\qquad-y^3+y=0\eqno{(6.1)}   $$
The second condition $(6.1)$ is satisfied for $y=0$, or $\pm1$, so that the first condition gives $b=0$, or $2b-6a=0$.  Since $b>0$, the crossing occurs at $b=3a$, so that the zeros of $P_{ab}$ remain in $\{z:{\rm Re}z<0\}$ for $b<3a$.
\medskip
	Since we have
$$	z^{-2}P_{ab}(z)=b(z^2+1/z^2)+4(z+1/z)+6a=b(z+1/z)^2+4(z+1/z)+6a-2b   $$
we may write $z^{-2}P_{ab}(z)=4R_{ab}(\zeta)$, with $2\zeta=z+1/z$ and
$$	R_{ab}(\zeta)=b\zeta^2+2\zeta+(3a-b)/2   $$
Since $R_{ab}(\zeta)=0$ corresponds to $\zeta=(-1\pm\sqrt{1-b(3a-b)/2}\,)/b$, assuming that $b\ge1$, and $2-b(3a-b)\ge0$, and $R_{ab}(-1)\ge0$, implies that the roots of $R_{ab}$ are real $\in[-1,1)$.  Therefore the roots of $P_{ab}$ are on the unit circle provided $b\ge1$, $3a\le b+2/b$ and $4\le3a+b$.
\medskip
	Assume $4\le3a+b$.  If $3a-b>0$, then (i) holds.  Otherwise $3a\le b$ which, together with $4\le3a+b$, implies $2\le b$, hence $1\le b$.  Also $3a\le b+2/b$ is implied by $3a\le b$, so that (ii) follows.  We have thus shown that $4\le3a+b$ implies (i) or (ii) (or both (i) and (ii)).
	
		Let
$$	g(J_2,J_4)=3-4e^{2\beta(J_2-J_4)}+e^{8\beta J_2}   $$
Then $g(0,0)=0$, and
$$	\partial_{J_2}g(J_2,0)=8\beta(e^{8\beta J_2}-e^{2\beta J_2})   $$
is $>0$ if $J_2>0$, and $<0$ if $J_2<0$.  Hence $g(J_2,0)\ge0$ for all $J_2$.  Since $g(J_2,J_4)\ge g(J_2,0)$ when $J_4\ge0$, we have $g(J_2,J_4)\ge0$ if $J_4\ge0$, for all $J_2$.
\bigskip
	{\bf 7. The zeros of the general partition function $Z_\Lambda$.}
\medskip
	Let $\Lambda$ be a union of distinguished squares $\alpha$.  As noted in Section 2, $Z_\Lambda$ is obtained from the product $\Pi$ of the partition functions $Z^\alpha$ by some relabelings $z_i^\alpha\to z_x$, and by Asano contractions $(z_i^\alpha,z_j^\beta)\to z_x$.  According to Proposition 6 we have $\Pi\ne0$ when $z_i^\alpha\notin K\subset{\bf C}$ with different choices (depending on $a,b$) for the closed set $K\not\ni0$.
	
\noindent
(i) If $3a-b>0$ we may choose $K=\{z:{\rm Re}z<-\epsilon\}$ for some $\epsilon>0$, and the complement of $-KK$ is a neighborhood of the positive real axis (the inside of a parabola).  Therefore $Z_\Lambda\ne0$ when all $z_x$ are close to the positive real axis.

\noindent
(ii)	If $1\le b$, $3a\le b+2/b$, and $4\le3a+b$, we may choose $K=\{z:|z|\ge1\}$, so that $-KK$ is again $\{z:|z|\ge1\}$.  Therefore $Z_\Lambda\ne0$ when all $z_x$ are in $\{z:|z|<1\}$, and also, by spin-flip symmetry, when all $z_x$ are in $\{z:|z|>1\}$.
\bigskip
	{\bf 8. GKS and FKG inequalities for multispin interactions.}
\medskip
	Consider the energy function
$$	H(\underline\sigma)=-\sum_{A\subset\Lambda}J_A\sigma^A   $$
where $\sigma^A=\prod_{x\in A}\sigma_x$.  We denote by $\langle\cdots\rangle$ the expectation value which gives the spin configuration $\underline\sigma$ the probability $\mu(\underline\sigma)=Z_\Lambda^{-1}\exp(-\beta H(\underline\sigma))$.
\medskip
	{\sl Assume $J_X\ge0$ for all $X\subset\Lambda$ (in particular $h_x=J_{\{x\}}\ge0$ for all $x\in\Lambda$).  The GKS inequalities then hold:
$$	\langle\sigma^A\rangle\ge0\qquad,\qquad
	\beta^{-1}\partial\langle\sigma^A\rangle/\partial J_B=
\langle\sigma^A\sigma^B\rangle-\langle\sigma^A\rangle\langle\sigma^B\rangle\ge0   $$
for all $A,B\subset\Lambda$.}
\medskip
	[See for instance [G] for a proof].
\medskip
	Let now
$$	H(\underline\sigma)
	=U(\underline\sigma)-\sum_{x\in\Lambda}h_x\sigma_x   $$
with $U=U'+U''$:
$$	U'(\underline\sigma)
	=-\sum_{A\subset\Lambda}K'_A\prod_{x\in A}{1+\sigma_x\over2}\quad,\quad
	U''(\underline\sigma)=-\sum_{A\subset\Lambda}K''_A
	\prod_{x\in A}{1-\sigma_x\over2}   $$
(We may take $K'_A,K''_A=0$ if $|A|=1$).  We write $\mu(\underline\sigma)=\mu_X$ for the probability associated with the spin configuration $\underline\sigma$ corresponding to the set $X=\{x:\sigma_x=+1\}$ of ``occupied'' sites.  Note that the spin-flip symmetric situation occurs when $K'_A=K''_A$ for all $A$ with $|A|>2$.  The $|A|=2$ case is always symmetric, aside from the one-body terms  which can be absorbed in the $\{h_x\}$.
\medskip
	{\sl Assume that $K'_A,K''_A\ge0$ for all $A\subset\Lambda$.  The FKG inequalities then hold, i.e., if $\underline\sigma\mapsto f(\underline\sigma),g(\underline\sigma)$ are nondecreasing functions, then $\mu(fg)\ge\mu(f)\mu(g)$} [FKJ].
\medskip	
	The proof is based on checking Holley's criteria for FKG [H]:
$$	\mu_{X\cup Y}\cdot\mu_{X\cap Y}\ge\mu_X\cdot\mu_Y   $$
for all $X,Y\subset\Lambda$.
\medskip
	The $h_x$ drop out of the formula.  Writing $U_X$ instead of $U(\underline\sigma)$ we have to check that
$$	-U_{X\cup Y}-U_{X\cap Y}\ge-U_X-U_Y   $$
for $U=U'+U''$, where $U'_X=-\sum_{A\subset X}K'_A$, and $U''_X=-\sum_{A\subset\Lambda\backslash X}K''_A$.  Define ${\bf1}_X(A)=1$ if $A\subset X$, $=0$ otherwise.  It suffices to prove
$$	\sum_AK'_A({\bf1}_{X\cup Y}(A)+{\bf1}_{X\cap Y}(A))
	\ge\sum_AK'_A({\bf1}_X(A)+{\bf1}_Y(A))   $$
and similarly with $X,Y$ replaced by $\Lambda\backslash X$, $\Lambda\backslash Y$, and $K'_A$ by $K''_A$.  Since $K'_A,K''_A\ge0$, it suffices to show that
$$	{\bf1}_{X\cup Y}(A)+{\bf1}_{X\cap Y}(A)\ge{\bf1}_X(A)+{\bf1}_Y(A)   $$
or
$$	{\bf1}_{X\cup Y}(A)-{\bf1}_X(A)\ge{\bf1}_Y(A)-{\bf1}_{X\cap Y}(A)   $$
This is true because the left-hand side can vanish only if $A\not\subset X\cup Y$ or if $A\subset X$, and this implies that the right-hand side also vanishes.
\medskip
	The combination of GKS and FKG inequalities , which both hold in the ferromagnetic examples considered, yield much information about the equilibrium properties of the system (see [L1], [L2]).  It shows in particular that the equilibrium states are linear combinations of those obtained from all-plus and all-minus boundary conditions when the free energy is differentiable in $\beta$.
\bigskip
	{\bf 9. Proof of the main results.}
\medskip
	Most of the main results stated in Section 3 have already been verified in Section 7.  We now have to check that a phase transition actually occurs at low temperature if $J_2>0,J_4\ge0$.  Putting $J_4$ and the diagonal interactions on distinguished squares $=0$, we recover, if $J_2>0$, the standard ferromagnetic Ising model: this has spontaneous magnetization at high $\beta$.  The GKS inequality (Section 8) shows that the spontaneous magnetization persists when diagonal interactions on distinguished squares are introduced, and a 4-spin interaction $J_4\ge0$ is added.  Finally, if the 4-spin interaction is actually present, i.e., $\Lambda$ contains at least one distinguished square, and $J_4\ne0$, the Lee-Yang property is actually violated at sufficiently high temperature, as proved in [R3] Theorem 9.
\medskip
	We shall now give a sufficient condition such that the Lee-Yang property holds at low temperature for multispin interactions.
\bigskip
	{\bf 10. Proposition.}
\medskip
	{\sl Define the partition function $Z_\Lambda(\underline z)$ associated with a finite set $\Lambda$ by
$$	Z_\Lambda(\underline z)=\sum_{X\subset\Lambda}[\exp(-\beta U_X)]z^X   $$
where
$$	U_X=U(\underline\sigma)=-\sum_{A\subset\Lambda}K_A
	(\prod_{x\in A}{1+\sigma_x\over2}+\prod_{x\in A}{1-\sigma_x\over2})
	=-(\sum_{A\subset X}K_A+\sum_{A\subset{\Lambda\backslash X}}K_A)   $$
so that we have the spin-flip symmetry $U_X=U_{\Lambda\backslash X}$.  Suppose that for each $A\subset\Lambda$ either $K_A=0$ or $\beta K_A\ge(|A|-1)\log2$, then the Lee-Yang property is satisfied: $Z_\Lambda(\underline z)\ne0$ if $|z_x|<1$ for all $x\in\Lambda$. [For $|A|=2,3$ it suffices to assume $\beta K_A\ge0$, and for $|A|=4$ it suffices to assume $\beta K_A\ge\log2$].}
\medskip
	[Note that this is quite general, and applies in particular to a version of the models in Sections 2 and 4 having 4-spin interactions on all squares.]
\medskip
	The Lee-Yang property is preserved by the product of multiaffine polynomials in disjoint sets of variables, and by Asano contraction (see Lemma A1 in the Appendix below).  Therefore, it suffices to prove the Lee-Yang property for the energy function
$$	U^{(A)}=-K_A(\prod_{x\in A}{1+\sigma_x\over2}+\prod_{x\in A}{1-\sigma_x\over2})   $$
defined on spin configurations in $A$.  We have thus to study the zeros of
$$	Z^{(A)}=\sum_{X\subset A}[\exp(-\beta U_X^{(A)})]z^X
	=\prod_{x\in A}(1+z_x)+(e^{\beta K_A}-1)(1+\prod_{x\in A}z_x)   $$
which is a symmetric function of the $z_x$, $x\in A$.  Using the characterization of Lee-Yang polynomials given in [R3] Theorem 3, and Grace's Theorem (Theorem A2 in the Appendix below), one finds (see [R3] Remark 4(a)) that $Z^{(A)}$ satisfies the Lee-Yang property provided
$$	(1+z)^{|A|-1}+(e^{\beta K_A}-1)   $$
does not vanish when $|z|<1$.  For $|A|>2$, the largest negative number of the form $(1+z)^{|A|-1}$, with $|z|\le1$, is obtained when ${\rm arg}(1+z)=\pi/(|A|-1)$, or $z=\exp(2i\pi/(|A|-1))$.  Thus, for $|z|\le1$,
$$	(1+z)^{|A|-1}\ge(1+\exp({2i\pi\over{|A|-1}}))^{|A|-1}>1-2^{|A|-1}   $$
Therefore the Lee-Yang property of $Z^{(A)}$ is ensured by $\exp(\beta K_A)-1\ge2^{|A|-1}-1$, i.e., $\beta K_A\ge(|A|-1)\log2$.  This was our assertion for general $|A|$. If $|A|=2$, it suffices to take $K_A\ge0$, because ferromagnetic pair interactions imply the Lee-Yang property.  The case $|A|=3$ reduces to the case $|A|=2$ because
$$	{1+\sigma_1\over2}\cdot{1+\sigma_2\over2}\cdot{1+\sigma_3\over2}
	+{1-\sigma_1\over2}\cdot{1-\sigma_2\over2}\cdot{1-\sigma_3\over2}
	={1\over4}[\sigma_1\sigma_2+\sigma_2\sigma_3+\sigma_3\sigma_1+1]   $$
$$	={1\over2}[({1+\sigma_1\over2}\cdot{1+\sigma_2\over2}
	+{1-\sigma_1\over2}\cdot{1-\sigma_2\over2})+{\rm permutations}-1]   $$
The case $|A|=4$ results from our analysis in Proposition 6.
\bigskip
	{\bf Acknowledgments.}
\medskip
	JLL thanks the IHES for its hospitality.  We are indebted to Joseph Slawny for helpful comments, and DR thanks J\"urg Fr\"ohlich for a useful conversation.  The work of JLL was supported in part by NSF grant DMR-0802120 and AFOSR grant AF-FA 9550-07.
\bigskip
	{\bf A. Appendix: Basic general results on multiaffine polynomials.}\footnote{(*)}{See the Appendix of [R3] for more details, and [BB1], [BB2] for a much more general approach, with many references.}
\medskip
	{\bf A1. Lemma} (Asano-Ruelle) [A], [R2].
\medskip
	{\sl Let $K_1,K_2$ be closed subsets of ${\bf C}$, with $K_1,K_2\not\ni0$.  If $\Phi$ is separately linear in $z_1$ and $z_2$, and if 
$$	\Phi(z_1,z_2)\equiv A+Bz_1+Cz_2+Dz_1z_2\ne0      $$
whenever $z_1\notin K_1$ and $z_2\notin K_2$, then
$$	\tilde\Phi(z)\equiv A+Dz\ne0      $$
whenever $z\notin-K_1\cdot K_2$.  {\rm [}We have written $-K_1\cdot K_2=\{-uv:u\in K_1,v\in K_2\}${\rm ]}.}
\medskip
	The map $\Phi\mapsto\tilde\Phi$ is called {\it Asano contraction}.
\medskip
	{\bf A2. Theorem} (Grace's theorem).
\medskip
	{\sl Let $P$ be a complex polynomial of degree $n$ in one variable and $\Phi$ be the only polynomial symmetric in its $n$ arguments, separately linear in each, such that
$$	\Phi(z,\ldots,z)=P(z)      $$
If the $n$ roots of $P$ are contained in a closed circular region $K$ and $z_1\notin K,\ldots,z_k\notin K$, then $\Phi(z_1,\ldots,z_n)\ne0$.}
\medskip
	A closed circular region is a closed subset $K$ of ${\bf C}$ bounded by a circle or a straight line.  We allow the coefficients of $z^n,z^{n-1},\ldots$ in $P$ to vanish: we then say that some  of the roots of $P$ are at $\infty$, and we take $K$ noncompact.
\medskip
	For a proof see Polya and Szeg\"o [PSz] V, Exercise 145.
\bigskip
	{\bf B. Appendix: model without diagonal interactions.}
\medskip
	The present model corresponds to the energy (5.1) in which the ``diagonal'' pairs $1<3$, $2<4$ have been removed from the sum over $i<j$.  The partition function of the system of 4 spins on a distinguished square, with vertices labeled consecutively 1,2,3,4, thus has the form
$$	Z(z_1,z_2,z_3,z_4)=e^{2\beta K_2}[e^{\beta(K_4+2K_2)}+(z_1+z_2+z_3+z_4)
	+(z_1z_2+z_2z_3+z_3z_4+z_4z_1)   $$
$$	+e^{-2\beta K_2}(z_1z_3+z_2z_4)
	+(z_1z_2z_3+z_2z_3z_4+z_3z_4z_1+z_4z_1z_2)
	+e^{\beta(K_4+2K_2)}z_1z_2z_3z_4]   $$
For this model we shall prove that if $K_2,K_4\ge0$, either or both of the following properties hold:

\noindent
(i) there is a neighborhood (independent of $\Lambda$) of the positive real axis which is free of zeros of $Z_\Lambda(z\ldots,z)$,

\noindent
(ii) $Z_\Lambda$ satisfies the Lee-Yang property.
\medskip
	Note that $Z(z_1,z_2,z_3,z_4)$ is separately symmetric in $z_1,z_3$, and $z_2,z_4$.  Using Grace's Theorem, we can analyze the zeros of $Z$ in terms of the zeros of
$$	P(u,v)=b+2(u+v)+4uv+a(u^2+v^2)+2(u+v)uv+bu^2v^2   $$
$$	=b(1+u^2v^2)+2(u+v)(1+uv)+4uv+a(u^2+v^2)   $$
where $a=e^{-2\beta K_2}$, $b=e^{\beta(K_4+2K_2)}$.  Note that the case $K_2=0$ has already been covered in Section 6, so that we can take $K_2>0$.  Also, since $K_4=0$ corresponds to the classical Ising model, we can take $K_4>0$.  We have thus to consider only the situations where $K_2,K_4>0$, i.e., $0<a<1$, and $ab>1$, which implies $a+b>a+1/a\ge2$.  In fact, the only assumptions we shall use in what follows are $0<a<1$ and $0<a+b-2$.
\medskip
	(i) We prove that if $|b-a|<2$, then $P(u,v)\ne0$ when ${\rm Re}\,u,{\rm Re}\,v\ge0$ (this requires only $a,b>0$).
\medskip
	For $a=b=1$, we have $	P(u,v)=(1+u)^2(1+v)^2$.  One can see that for the zeros $(u,v)$ of $P$ to reach the region ${\rm Re}\,u,{\rm Re}\,v\ge0$, they must intersect ${\rm Re}\,u={\rm Re}\,v=0$.  We have (with real $x,y$):
$$	P(ix,iy)=b(1+x^2y^2)+2i(x+y)(1-xy)-4xy-a(x^2+y^2)   $$
and $P(ix,iy)=0$ yields either (a) or (b):

(a): $x+y=0$ and $b(1+x^4)+(4-2a)x^2=0$, or $x^4-2({a-2\over b})x^2+1=0$, hence ${a-2\over b}\ge1$, i.e., $a-b\ge2$,

(b): $xy=1$ and $2b-4-a(x^2+1/x^2)=0$, or $x^4-2({b-2\over a})x^2+1=0$, hence ${b-2\over a}\ge1$, i.e., $b-a\ge2$.

Therefore, if $|b-a|<2$ and $P(u,v)=0$, we cannot have ${\rm Re}\,u\ge0\,\,{\rm and}\,\,{\rm Re}\,v\ge0$.  There is thus $\epsilon>0$ such that, writing $L=\{z:{\rm Re}\,z\le-\epsilon\}$, we have
$$	z_1,z_2,z_3,z_4\not\in L\qquad{\rm implies}\qquad Z(z_1,z_2,z_3,z_4)\ne0   $$
\medskip
	(ii) The Lee-Yang property for $Z$ is equivalent to
$$	b+(z_1+z_2+z_3)+(z_1z_2+z_2z_3)+az_1z_3+z_1z_2z_3\ne0
	\qquad{\rm if}\qquad|z_1|,|z_2|,|z_3|<1   $$
by [R3] Theorem 3, or to
$$	u(1+v)^2+b+2v+av^2\ne0\qquad{\rm if}\qquad|u|,|v|<1\eqno{(B.1)}   $$
by Grace's Theorem.  Since $a+b-2>0$, we may define $\lambda=(ab-1)/(a+b-2)$, so that $(1-\lambda)^2=(b-\lambda)(a-\lambda)$ and $a-\lambda=(1-a)^2/(a+b-2)>0$.  Since $1-a>0$, the above condition $(B.1)$ may now be rewritten as 
$$	(u+\lambda)(1+v)^2+(a-\lambda)(v^2+2{1-\lambda\over a-\lambda}v
	+({1-\lambda\over a-\lambda})^2)\ne0   $$
or
$$	u+\lambda+(a-\lambda)\Big({v+{1-\lambda\over a-\lambda}\over v+1}\Big)^2\ne0
	\qquad{\rm for}\qquad|u|,|v|<1   $$
which is equivalent to
$$	u+\lambda+(a-\lambda)w^2\ne0\qquad{\rm for}\qquad|u|<1,{\rm Re}\,w>\big({1\over2}(1+{1-\lambda\over a-\lambda})\big)^2\eqno{(B.2)}   $$
because $v\mapsto(v+{1-\lambda\over a-\lambda})/(v+1)$ maps the disk $\{v:|v|<1\}$ to the half-plane $\{w:{\rm Re}\,w>{1\over2}(1+{1-\lambda\over a-\lambda})\}$.
\medskip
	The boundary of $\{w^2:{\rm Re}\,w>{1\over2}(1+{1-\lambda\over a-\lambda})\}$ is a parabola $\{[{1\over2}(1+{1-\lambda\over a-\lambda})+it]^2:t\in{\bf R}\}$.  Therefore we want
$$	u+\lambda+(a-\lambda)[{1\over2}(1+{1-\lambda\over a-\lambda})+it]^2\ne0   $$
or
$$	u+\lambda+{1\over4}(a-\lambda)+{1\over2}(1-\lambda)+{1\over4}(b-\lambda)
	+(a-\lambda+1-\lambda)it-(a-\lambda)t^2\ne0   $$
or
$$	u+{1\over4}(a+b+2)+(1+a-2\lambda)it-(a-\lambda)t^2\ne0
	\qquad{\rm for}\qquad|u|<1,t\in{\bf R}\eqno{(B.3)}   $$
Since $a+b+2>0$, one checks that $(B.2)$ is equivalent to $(B.3)$.
\medskip
	We rewrite $(B.3)$ as
$$	u+C+Bit-At^2\ne0\qquad{\rm if}\qquad|u|<1,t\in{\bf R}\eqno{(B.4)}   $$
with
$$	A=a-\lambda={(1-a)^2\over a+b-2}   $$
$$	B=1+a-2\lambda={(1-a)(b-a)\over a+b-2}   $$
$$ C={1\over4}(a+b+2)   $$
$(B.4)$ means that the distance of $0$ to the parabola $C+Bit-At^2$ is $\ge1$.  For the closest point to $0$ of the parabola, we have $-2At(C-At^2)+B^2t=0$, i.e., $t=0$, or $t^2=C/A-B^2/2A^2$ if this quantity is $>0$.  This gives either $dist=C>1$, because $a+b>2$, or if $t^2=C/A-B^2/2A^2>0$:
$$ dist^2=(C-A({C\over A}-{B^2\over2A^2}))^2+B^2({C\over A}-{B^2\over2A^2})
={B^4\over4A^2}+B^2({C\over A}-{B^2\over2A^2})=B^2({C\over A}-{B^2\over4A^2})   $$
In this last case $dist^2\ge1$ is equivalent to
$$	{B^4\over4A^2}+1\le{C\over A}B^2   $$
but since we have $C/A-B^2/2A^2>0$, hence $(C/A)B^2>B^4/2A^2$, it suffices to check that $B^4/4A^2+1\le B^4/2A^4$, i.e., $1\le B^4/4A^2$, or
$$	1\le{B^2\over2A}={(b-a)^2\over2(a+b-2)}   $$
or $(b-a)^2\ge2(a+b-2)$, or $(b-a-1)^2\ge4a-3$.
\medskip
	We have thus proved that the Lee-Yang property holds under the assumptions that $0<a<1,\,0<a+b-2$, and
$$	(b-a-1)^2\ge4a-3   $$
but either (i) holds, or $b-a\ge2$ , i.e., $b-a-1\ge1$, hence $(b-a-1)^2\ge4a-3$.  We have thus shown that, under the assumptions $K_2>0$, $K_4>0$, which imply $0<a<1$, $0<a+b-2$, we have either (i) or (ii) or both, as announced.
\bigskip
	{\bf References.}
\medskip
[A] T. Asano.  ``Theorems on the partition functions of the Heisenberg ferromagnets.'' J. Phys. Soc. Jap. {\bf 29},350-359(1970).

[BB1] J. Borcea and P. Br\"and\'en.  ``The Lee-Yang and Polya-Schur programs. I. Linear operators preserving stability.'' Inventiones Math. {\bf 177},541-569(2009).

[BB2] J. Borcea and P. Br\"and\'en.  ``The Lee-Yang and Polya-Schur programs. II. Theory of stable polynomials and applications.''  Commun. Pure Appl. Math. {\bf 62},1595-1631(2009).

[FKG] C.M. Fortuin, P.W. Kasteleyn, and J. Ginibre.  ``Correlation inequalities on some partially ordered sets.''  Commun. Math. Phys. {\bf 22}, 89-103(1971).

[G] J. Ginibre.  ``General formulation of the Griffiths' inequalities.''  Commun. Math. Phys. {\bf 16},310-328(1970).

[H] R. Holley.  ``Remarks on the FKG inequalities.''  Commun. Math. Phys. {\bf 36},227-231(1974).

[HS] W. Holsztynski and W. Slawny.  ``Phase transitions in ferromagnetic spin systems at low temperatures.''  Commun. Math. Phys. {\bf 66},147-166(1979).

[I] R.B. Israel.  {\it Convexity in the theory of lattice gases.}  Princeton U.P., Princeton, 1979.

[L1] J.L. Lebowitz.   ``GHS and other inequalities.''  Commun. Math. Phys. {\bf 35},87-92(1974).

[L2] J.L. Lebowitz.   ``Coexistence of phases in Ising ferromagnets.''  J. Statist. Phys. {\bf 16},463-476(1977).

[LR]  E.H. Lieb and D. Ruelle.  ``A property of zeros of the partition function for Ising spin systems.''  J. Math. Phys. {\bf13},781-784(1972).

[LY] T. D. Lee and C. N. Yang.  ``Statistical theory of equations of state and phase relations. II.  Lattice gas and Ising model.''  Phys. Rev. {\bf 87},410-419(1952).

[PSz] G. Polya and G. Szeg\"o.  {\it Problems and theorems in analysis II.}  Springer, Berlin, 1976.

[R1] D. Ruelle.  {\it Statistical Mechanics.  Rigorous Results.}  Benjamin, New York, 1969.  (Reprint: Imperial College Press, London, and World Scientific, Singapore, 1999).

[R2] D. Ruelle.  ``Extension of the Lee-Yang circle theorem.''  Phys. Rev.
Letters {\bf 26},303-304(1971).

[R3] D. Ruelle.  ``Characterization of Lee-Yang polynomials.''  Ann. of Math. {\bf 171},589-603(2010).

[Sim] B. Simon.  {\it The Statistical Mechanics of Lattice Gases.} Vol. I, Princeton U.P., Princeton, 1993.

[Sin] Ya.G. Sinai.  {\it Theory of Phase Transitions: Rigorous Results.}  Pergamon, Oxford, 1982.

[Sl] J. Slawny.  ``Low-temperature properties of classical lattice systems: phase transitions and phase diagrams'' in {\it Phase Transitions and Critical Phenomena}. Vol. {\bf 11}, C. Domb and J.L. Lebowitz, eds., Academic Press, London, 1987.
\end